\documentclass[10pt,a4paper,twocolumn,superscriptaddress,aps,prd,longbibliography,nofootinbib]{revtex4-2}

\usepackage{lineno} 

\usepackage{amsmath}
\usepackage{amsfonts}
\usepackage{amssymb}
\usepackage{bbold}
\usepackage{graphicx}
\usepackage[usenames,dvipsnames]{color}
\usepackage{float}
\usepackage{multirow}
\usepackage{soul}
\usepackage{siunitx}
\usepackage{placeins}
\usepackage[nohyperlinks, printonlyused, nolist]{acronym}
\usepackage {xcolor}

\usepackage{blindtext}

\begin{document}

\setlength\linenumbersep{0.18cm}

\author{Lukas Gehrig} 
\affiliation{Physikalisches Institut, Universit\"at W\"urzburg, D-97074 W\"urzburg, Germany}
\affiliation{W\"urzburg-Dresden Cluster of Excellence ct.qmat, Universit\"at W\"urzburg, D-97074 W\"urzburg, Germany} 

\author{Cedric Schmitt} 
\affiliation{Physikalisches Institut, Universit\"at W\"urzburg, D-97074 W\"urzburg, Germany}
\affiliation{W\"urzburg-Dresden Cluster of Excellence ct.qmat, Universit\"at W\"urzburg, D-97074 W\"urzburg, Germany} 

\author{Jonas Erhardt} 
\affiliation{Physikalisches Institut, Universit\"at W\"urzburg, D-97074 W\"urzburg, Germany}
\affiliation{W\"urzburg-Dresden Cluster of Excellence ct.qmat, Universit\"at W\"urzburg, D-97074 W\"urzburg, Germany} 

\author{Bing Liu} 
\affiliation{Physikalisches Institut, Universit\"at W\"urzburg, D-97074 W\"urzburg, Germany}
\affiliation{W\"urzburg-Dresden Cluster of Excellence ct.qmat, Universit\"at W\"urzburg, D-97074 W\"urzburg, Germany} 

\author{Tim Wagner}
\affiliation{Physikalisches Institut, Universit\"at W\"urzburg, D-97074 W\"urzburg, Germany}
\affiliation{W\"urzburg-Dresden Cluster of Excellence ct.qmat, Universit\"at W\"urzburg, D-97074 W\"urzburg, Germany}

\author{Martin Kamp}
\affiliation{Physikalisches Institut, Universit\"at W\"urzburg, D-97074 W\"urzburg, Germany}
\affiliation{Physikalisches Institut and R\"ontgen Center for Complex Material Systems, D-97074 W\"urzburg, Germany}

\author{Simon Moser}
\affiliation{Physikalisches Institut, Universit\"at W\"urzburg, D-97074 W\"urzburg, Germany}
\affiliation{W\"urzburg-Dresden Cluster of Excellence ct.qmat, Universit\"at W\"urzburg, D-97074 W\"urzburg, Germany}

\author{Ralph Claessen}
\email{e-mail: claessen@physik.uni-wuerzburg.de}
\affiliation{Physikalisches Institut, Universit\"at W\"urzburg, D-97074 W\"urzburg, Germany}
\affiliation{W\"urzburg-Dresden Cluster of Excellence ct.qmat, Universit\"at W\"urzburg, D-97074 W\"urzburg, Germany}

\date{\today}


\title{
Graphene intercalation of the large gap quantum spin Hall insulator
bismuthene\\
}

\maketitle


\bigskip\noindent{\bf 
The quantum spin Hall insulator bismuthene, a two-third monolayer of bismuth on SiC(0001), is distinguished by helical metallic edge states that are protected by a groundbreaking $800$\;meV topological gap, making it ideal for room temperature applications. This massive gap inversion arises from a unique synergy between flat honeycomb structure, strong spin orbit coupling, and an orbital filtering effect that is mediated by the substrate. However, the rapid oxidation of bismuthene in air has severely hindered the development of applications, so far confining experiments to ultra-high vacuum conditions. Here, we successfully overcome this barrier, intercalating bismuthene between SiC and a protective sheet of graphene. As we demonstrate through scanning tunneling microscopy and photoemission spectroscopy, graphene intercalation preserves the structural and topological integrity of bismuthene, while effectively shielding it from oxidation in air. We identify hydrogen as the critical component that was missing in previous bismuth intercalation attempts. Our findings facilitate ex-situ experiments and pave the way for the development of bismuthene based devices, signaling a significant step forward in the development of next-generation technologies.
}
\bigskip

\begin{figure}[b!]
\begin{center}
\includegraphics[width=\linewidth]{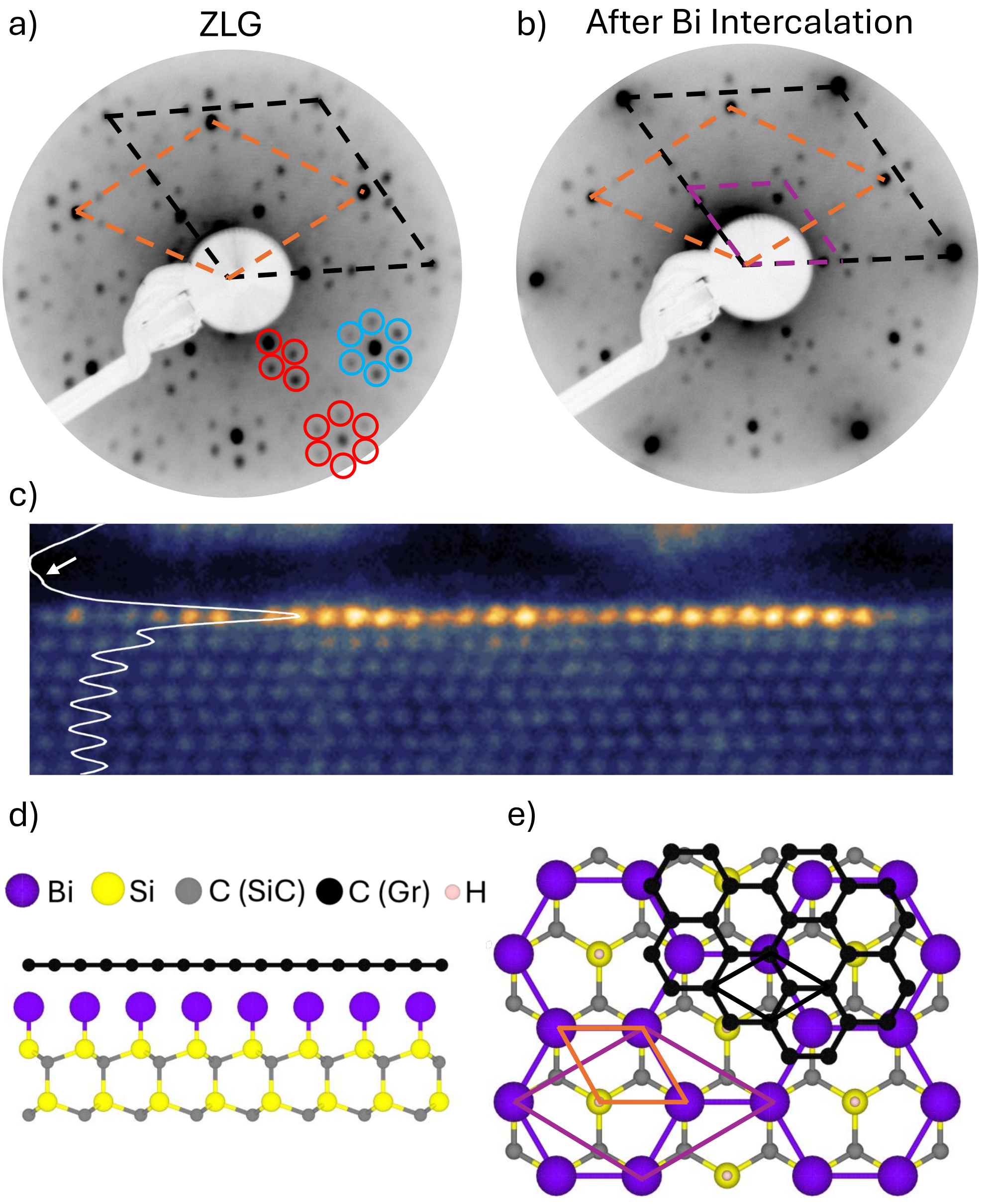}
\caption[Caption]{\textbf{Structural characterization of intercalated bismuthene}. LEED taken at 60\;eV before \textbf{a)} and after \textbf{b)} bismuthene intercalation. The spots marked by red and blue circles correspond to the $(6\sqrt{3} \times 6\sqrt{3})R30^\circ$ and quasi-$(6 \times 6)$ periodicity of ZLG, respectively. \textbf{c)} Cross-sectional TEM of intercalated bismuthene shows an atomically thin layer of bismuth between SiC and graphene. A shoulder (white arrow) in the horizontally integrated line profile (white line) shows the presence of the graphene sheet. 
Side view \textbf{d)} and top view \textbf{e)} of the ball and stick model of intercalated bismuthene. The black, orange and purple contours denote the primitive unit cells of graphene, SiC and bismuthene, respectively. By comparing the unit cells, it is evident that bismuthene is a $(\sqrt{3} \times \sqrt{3})R30^\circ$ superstructure with respect to SiC.}
\label{fig: VESTA_LEED_and_TEM}
\end{center}
\end{figure}

\begin{figure*}[t!!!!!!!]
\begin{center}
\includegraphics[width=\linewidth]{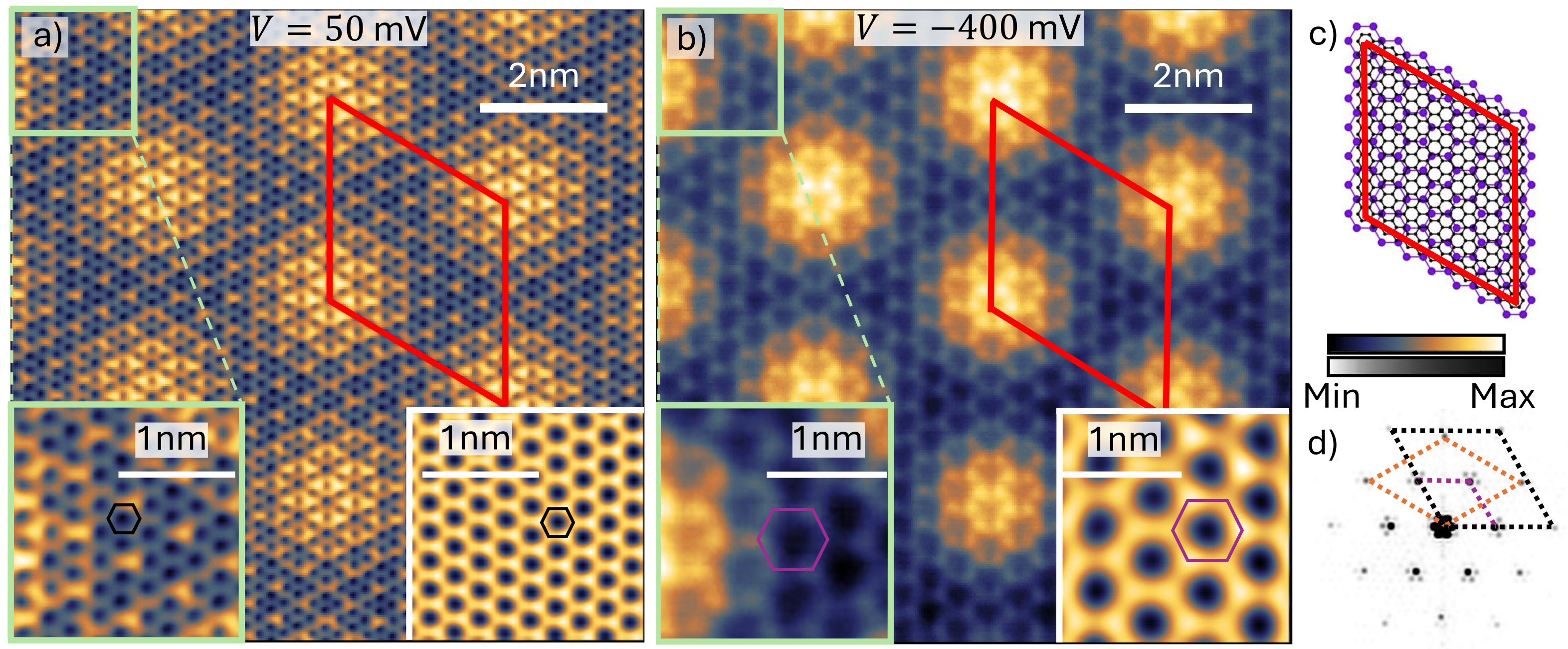}
\caption{\textbf{STM of intercalated bismuthene}. Constant current STM images measured with bias voltages and tunneling currents of \textbf{a)} $50$\,mV, $200$\,pA and \textbf{b)} $-400$\,mV, $180$\,pA, optimized to resolve the graphene and bismuthene lattices, respectively. Zoom-ins to the data are shown in the left insets. Corresponding Fourier filtered images are shown in the right inset. \textbf{c)} Moiré unit cell arising from the different periodicities of graphene and bismuthene. \textbf{d)} Fast Fourier Transform of the image in \textbf{b)}. The spots corresponding to graphene/SiC/bismuthene are linked by black/orange/purple dashed lines to mark the associated Brillouin zones, respectively.}
\label{fig: STM}
\end{center}
\end{figure*} 


Quantum spin Hall insulators (QSHIs) represent a groundbreaking class of quantum materials, characterized by the quantum spin Hall effect (QSHE) \cite{KaneMeleQSHE,Koenig2007}. These two-dimensional (2D) materials have garnered significant interest due to their spin-momentum locked metallic edge states \cite{Bernevig_QSHE}, which enable dissipationless electron transport. 
Such a unique property has profound implications for novel electronic \cite{Michetti_TI_Devices,Xiaofeng_QSHEinTMDC} and spintronic devices \cite{Liu2014,Han2018}. Additionally, the edge states of QSHIs may host Majorana fermions, positioning these materials as promising candidates for quantum computing \cite{Aasen2016_MajoranQC,Zhang_ReviewTI,Kane_ReviewTI}.


For QSHIs to serve in viable device applications, the spin-orbit inverted topological band gap $E_{\text{gap}}$ must substantially surpass thermal broadening, on the order of $E_{\text{gap}}\geq 3.5\; k_BT$, implying approximately 90\;meV at room-temperature (RT) \cite{Weber_2021}. This requirement excludes graphene, the material initially predicted to exhibit the QSHE \cite{KaneMeleQSHE}, due to its minuscule topological band gap of less than \SI{50}{\micro\electronvolt} \cite{GapGraphene}. Building on the honeycomb motif, researchers have substituted carbon atoms in graphene-like structures with heavier elements, thereby drastically enhancing spin-orbit coupling (SOC) and increasing the gap \cite{GangLi_Bismuthene}. A notable breakthrough in this area of 2D ``Xene''-materials is the discovery of bismuthene, a structural twin of graphene, realized as a 2/3 monolayer of bismuth on SiC(0001) that forms a $(\sqrt{3} \times \sqrt{3})R30^\circ$ structure \cite{ReisBismutheneScience}. Bismuthene's remarkable topological band gap of $E_{\text{gap}}\sim800$\,meV still remains the largest reported for any 2D topological insulator to date.


Despite this success, bismuthene's susceptibility to oxidation under ambient conditions poses a significant challenge, limiting its experimental use to ultra-high vacuum (UHV) environments. To mitigate this problem, metal intercalation underneath a protective sheet of graphene \cite{Riedl_H_Intercalation,Forti_Au_Intercalation,Lin_Sb_Intercalation,Matta_Pb_Intercalation,Rosenzweig_Ag_Intercalation} may prevent oxidation while preserving the material's topological properties, as we recently demonstrated for the QSHI indenene, i.e., a \textit{full} atomic monolayer of indium on SiC(0001) with a 120\;meV gap \cite{CedricSchmittIntercalation}. 

Although bismuth intercalation has been attempted, conventional methods have thus far yielded only trivial bismuth films. These include a fully covered yet metallic structure and a 1/3-covered insulating structure that remains topologically trivial \cite{Wolff_2024,Stöhr_Bi_Intercalation,Sohn_Bi_Intercalation}. Achieving intercalation of the topologically nontrivial 2/3 monolayer bismuthene phase, however, has proven more technologically challenging and has remained elusive.



In this work, we address this challenge and introduce a robust protocol for controlling the intercalation process between 1/3 and the full bismuth monolayers, and successfully stabilize the 2/3 monolayer QSHI bismuthene phase beneath a single graphene sheet. We demonstrate that this graphene layer effectively protects the intercalant from oxidation when exposed to controlled amounts of oxygen and air, while fully preserving its topologically non-trivial character. 



\begin{figure*}[t!!!!!!!!!!]
\begin{center}
\includegraphics[width=\linewidth]{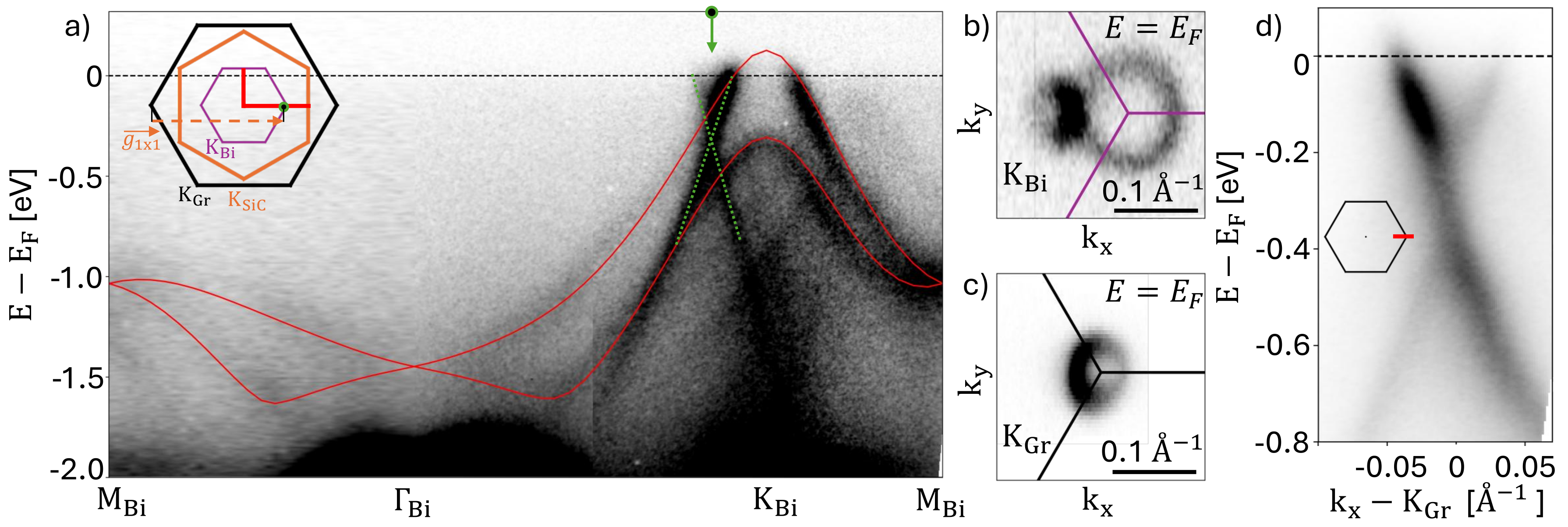}
\caption{\textbf{Band structure of intercalated bismuthene}. \textbf{a)} $M_\text{Bi} \Gamma K_\text{Bi} M_\text{Bi}$ high-symmetry band dispersion of intercalated bismuthene measured at room temperature with $21.2 \, \mathrm{eV}$ (He I$\alpha$) photon energy. A graphene replica due to photoelectrons scattering off the $(1 \times 1)$ lattice potential of the SiC substrate is also visible (green dashed lines). Fermi surfaces around the K-Points of bismuthene and graphene are shown in \textbf{b)} and \textbf{c)}. Panel \textbf{d)} highlights the Dirac crossing of the graphene $\pi$-bands.}
\label{fig: ARPES}
\end{center}
\end{figure*}


To produce bismuthene intercalated graphene, nitrogen-doped 4H-SiC substrates were first dry-etched in a hydrogen atmosphere, following the procedure outlined by Glass et al. \cite{Glass2016}. Subsequently, the substrates were annealed at 1100°C in UHV to epitaxially grow zero-layer graphene (ZLG), as previously reported by Riedl et al. in Ref. \onlinecite{Riedl_GrapheneOnSiC}. The resulting low-energy electron diffraction (LEED) pattern, shown in Fig. \ref{fig: VESTA_LEED_and_TEM}a), reveals the primitive periodicities of both the SiC(0001) surface and graphene, with their respective reciprocal unit cells outlined by orange and black contours. Furthermore, additional LEED peaks, highlighted by red and blue circles, correspond to the characteristic $(6\sqrt{3} \times 6\sqrt{3})R30^\circ$ and quasi-$(6 \times 6)$ periodicity associated with the ZLG superstructure \cite{Riedl_GrapheneOnSiC}.


After safe UHV transfer to our molecular beam epitaxy (MBE) system (Octoplus 300, Dr. Eberl MBE-Komponenten GmbH), bismuth (99.9999\% purity) was deposited from a Knudsen cell, followed by intercalation via annealing at 350°C. A subsequent bismuth desorption step at 550°C in hydrogen atmosphere to saturate the 1/3 silicon dangling bonds was critical to yield the desired bismuthene phase of 2/3 monolayer coverage. The corresponding LEED pattern in Fig. \ref{fig: VESTA_LEED_and_TEM}b) now shows a marked increase in the intensity of the primitive graphene spots, compared to the ZLG phase in Fig. \ref{fig: VESTA_LEED_and_TEM}a). This indicates the decoupling of ZLG from the SiC substrate and the formation of quasi-freestanding monolayer graphene (QFMG), in analogy to observations in previous graphene intercalation studies \cite{Bocquet_SbIntercalation,CedricSchmittIntercalation}. 


In addition, the emergence of new LEED spots after intercalation signifies the formation of a $(\sqrt{3} \times \sqrt{3})R30^\circ$ bismuth superstructure outlined in purple. Cross-sectional scanning transmission electron microscopy (STEM) images, shown in Fig. \ref{fig: VESTA_LEED_and_TEM}c), reveal the intercalated bismuth atoms to be indeed confined to a single atomic sheet, while the graphene layer is barely visible in the shoulder of the integrated line profile (white line) due to the low contrast of carbon compared to the heavy bismuth atoms \cite{CedricSchmittIntercalation,Briggs2020}. The STEM images also confirm that the intercalated bismuth atoms adsorb right atop the silicon atoms, i.e., at the T1 position of the SiC(0001) substrate. 

The combined LEED and STEM observations are consistent with a 2/3 monolayer bismuth phase with bismuth occupying the the T1 site directly above Si of the first SiC(0001) layer, as reported by Reis et al. \cite{ReisBismutheneScience}, but not with the reported 1/3 monolayer phase of previous studies, where bismuth is assumed to adsorb between the Si atoms  \cite{Wolff_2024,Stöhr_Bi_Intercalation, Sohn_Bi_Intercalation}. 
Therefore, we conclude the successful intercalation of bismuthene, proposing a structural model as presented in Fig. \ref{fig: VESTA_LEED_and_TEM}d) and e).


To determine the relative alignment and interaction between bismuthene and graphene, we present constant current scanning tunneling microscopy (STM) data of a representative $10\times 10$ nm$^2$ sample region in Fig. \ref{fig: STM}. Image a) was measured at a bias voltage of $50 \, \mathrm{mV}$, where STM is mostly susceptible to the honeycomb lattice of the graphene overlayer. In contrast, image b) was measured at $-400 \, \mathrm{mV}$ bias voltage, where STM probes deep into the bismuthene bulk bands as we confirm later by ARPES. Via the bias voltage, we thus selectively uncover the graphene overlayer as well as the bismuthene intercalant, as seen more clearly in the magnified insets on the left \cite{CedricSchmittIntercalation}. 
Further, both scans reveal the periodic Moiré pattern formed by the incommensurate bismuthene/SiC(0001) and graphene systems, whose lattice constant of approximately $a_\text{Moiré}=3.225 \, \mathrm{nm}$ corresponds to roughly $6 \, a_\text{Bi}$ or $13 \, a_\text{Gr}$. 


A fast Fourier transform (FFT) of image b) is shown in Fig. \ref{fig: STM}d) and further corroborates this analysis. Similar to the LEED pattern of intercalated bismuthene in Fig. \ref{fig: VESTA_LEED_and_TEM}b), the FFT reveals diffraction spots corresponding, respectively, to the SiC, the graphene, and the bismuthene lattices, marked by orange, black, and purple reciprocal unit cells. By selectively filtering the raw STM data for the spatial frequencies of each lattice, we obtain the clean graphene and bismuthene signatures shown in the right insets, yielding a Moiré unit cell according to Fig. \ref{fig: STM}c).

\begin{figure*}[t!!!!!]
\begin{center}
\includegraphics[width=\linewidth]{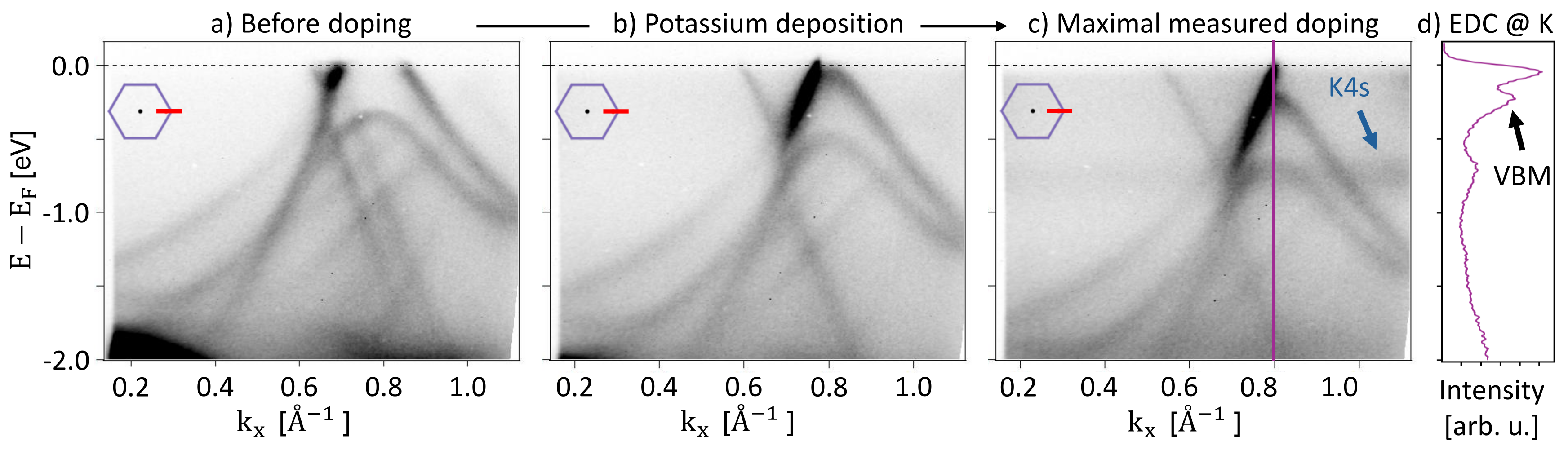}
\caption{\textbf{Potassium doping of intercalated bismuthene.} \textbf{a)} - \textbf{c)} ARPES at bismuthene K point for different levels of potassium doping, measured at $50 \, \mathrm{K}$ with a photon energy of $21.2 \, \mathrm{eV}$ (He I$\alpha$). \textbf{d)} Doping saturates with bismuthene's valence band maximum (VBM) at about $-230 \, \mathrm{meV}$, showing intercalated bismuthene to conserve its topological gap. A non-dispersive band observed at approximately $-760 \, \mathrm{meV}$ panel c) (blue arrow) is attributed to the $4s$ electrons of potassium.}
\label{fig: Doping}
\end{center}
\end{figure*}

With the structure of intercalated bismuthene now established, we turn our attention to its electronic properties. Room temperature angle-resolved photoemission spectroscopy (ARPES) measured with He I$\alpha$ radiation along the $M_\text{Bi} \Gamma K_\text{Bi} M_\text{Bi}$ high-symmetry path of bismuthene are presented in Fig. \ref{fig: ARPES}a). The distinctive $p_x/p_y$ low-energy band structure, characteristic of pristine bismuthene \cite{ReisBismutheneScience}, is clearly observed and matches the density functional theory (DFT) calculations (red lines) for the pristine phase \cite{GangLi_Bismuthene}. While a pocket in the Fermi surface of Fig. \ref{fig: ARPES}b indicates significant hole doping of $(9.2 \pm 2.4) \times 10^{12} \, \mathrm{cm}^{-2}$, the presence of a Rashba band splitting at $K_\text{Bi}$ suggests graphene and bismuthene to remain electronically decoupled. In contrast, the characteristic Dirac crossing of graphene's $\pi$-bands shown in Fig. \ref{fig: ARPES}d) is shifted to about -300 meV, indicating significant electron doping of $(4.4 \pm 1.9) \times 10^{12} \, \mathrm{cm}^{-2}$ extracted from graphene's Fermi surface in Fig. \ref{fig: ARPES}c). While this electron doping accounts only for about half of bismuthene's holes, the remaining hole concentration is likely contributed by the substrate, similar to the case of indenene \cite{CedricSchmittIntercalation}. 

Graphene's characteristic \textit{horseshoe} and \textit{dark corridor} features, arising from the honeycomb lattice structure factor, are shown in Fig. \ref{fig: ARPES}c) \cite{Bostwick2007_Graphene,Gierz2011_horseshoe,Moser_horseshoe}. Upon closer inspection, we observe a flipped replica of this horseshoe close to $K_\text{Bi}$ in Fig. \ref{fig: ARPES}b), producing the linear bands (green dashed lines) in the ARPES bandstructure of Fig. \ref{fig: ARPES}a). As sketched in the inset, this Umklapp arises from graphene photoelectrons scattering off the $(1 \times 1)$ lattice potential of the SiC substrate \cite{Origin_GrapheneReplica}, as indicated by the Umklapp vector $\vec{g}_{1\times 1}$ in Fig. \ref{fig: ARPES}a).

As the hole doping of bismuthene prevents us from an extraction of its topological gap, we use n-doping via deposition of potassium while monitoring the valence band around $K_{\text{Bi}}$ by ARPES \cite{CedricSchmittIntercalation}. As shown in Fig. \ref{fig: Doping}, this leads to a gradual filling of graphene and bismuthene with electrons, fully recovering bismuthene's valence band until its maximum saturates at about $230 \, \mathrm{meV}$ below the Fermi energy in Fig. \ref{fig: Doping}c). Although this doping experiment is incapable to resolve the bismuthene band gap in its entirety, the valence band saturation still provides an experimental lower limit of $E_{\text{gap}}\geq 230$\;meV, still sufficiently large as compared to thermal broadening effects at room temperature \cite{Weber_2021}, and likely just an extremely conservative understatement of the real bulk gap $E_{\text{gap}}\sim800$\;meV.

\begin{figure}[t!]
\begin{center}
\includegraphics[width=\linewidth]{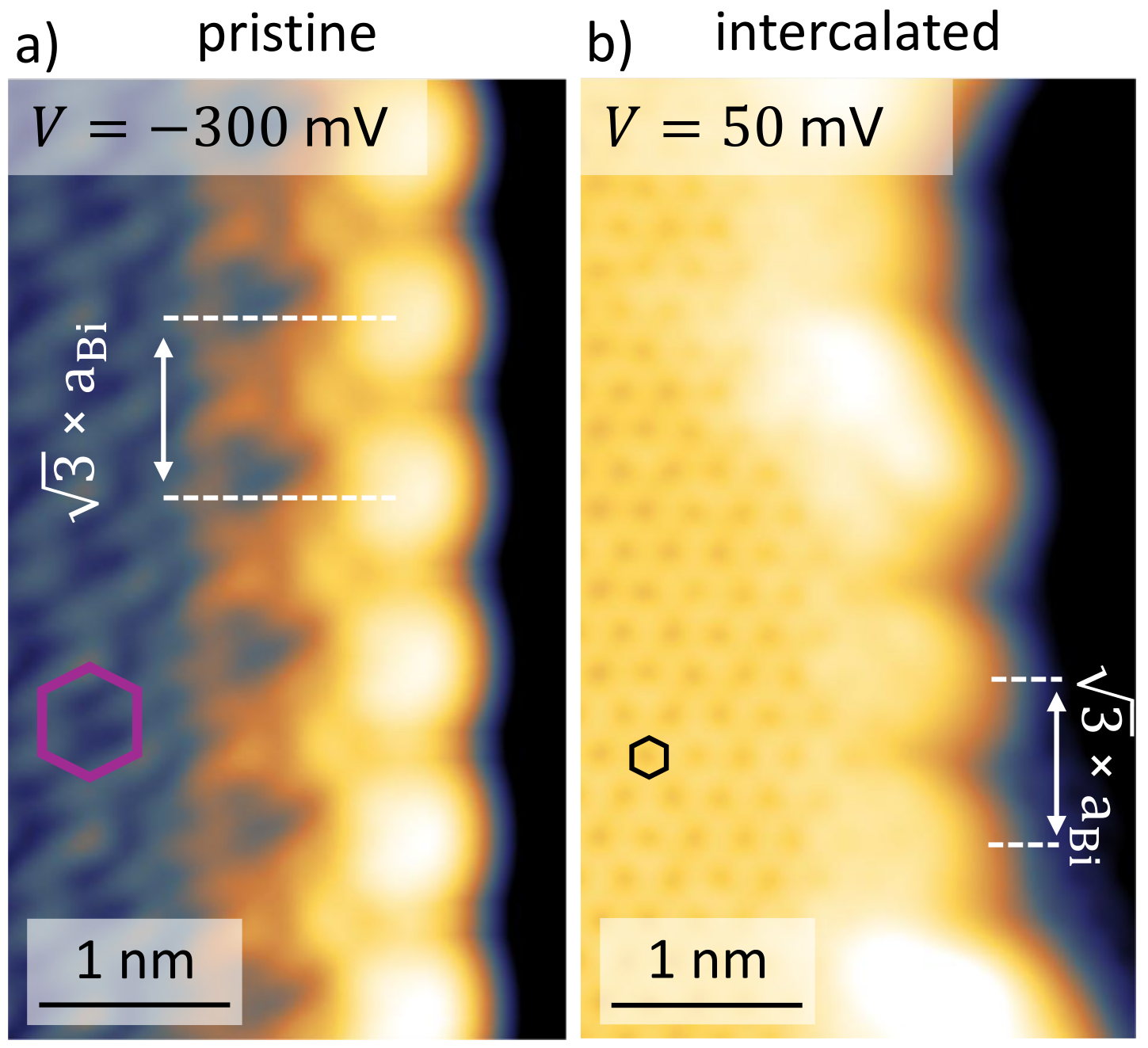}
\caption{\textbf{a)} Constant current STM images taken at bias voltage of $-300 \, \mathrm{mV}$ at a SiC step edge shows modulation, arising from topological edge states of pristine bismuthene. \textbf{b)} Intercalated bismuthene shows similar modulation at a bias voltage of $50 \, \mathrm{mV}$ at a SiC step edge.  This bias voltage corresponds to the same energy position as in a), accounting for the effective p-doping of the intercalated variant. A purple and a black hexagon marks the honeycomb of bismuthene and graphene, respectively.}
\label{fig: Edge}
\end{center}
\end{figure}

As a consequence of the topological bulk-boundary correspondence, pristine bismuthene displays metallic edge states filling the substantial band gap. The STM image of Fig.~\ref{fig: Edge}a) displays such a situation at the armchair termination of the bismuthene film at a SiC terrace step, reflected by the enhanced tunneling signal and a characteristic $\sqrt{3}$ periodicity \cite{Stuhler2020}. Quite remarkably, we observe identical behavior (enhanced signal and $\sqrt{3}$ modulation) in the armchair-edges of intercalated bismuthene (see Fig.~\ref{fig: Edge}b), strongly suggesting the conservation of the topological edge states despite the graphene overlayer. 

\begin{figure}[t!]
\begin{center}
\includegraphics[width=\linewidth]{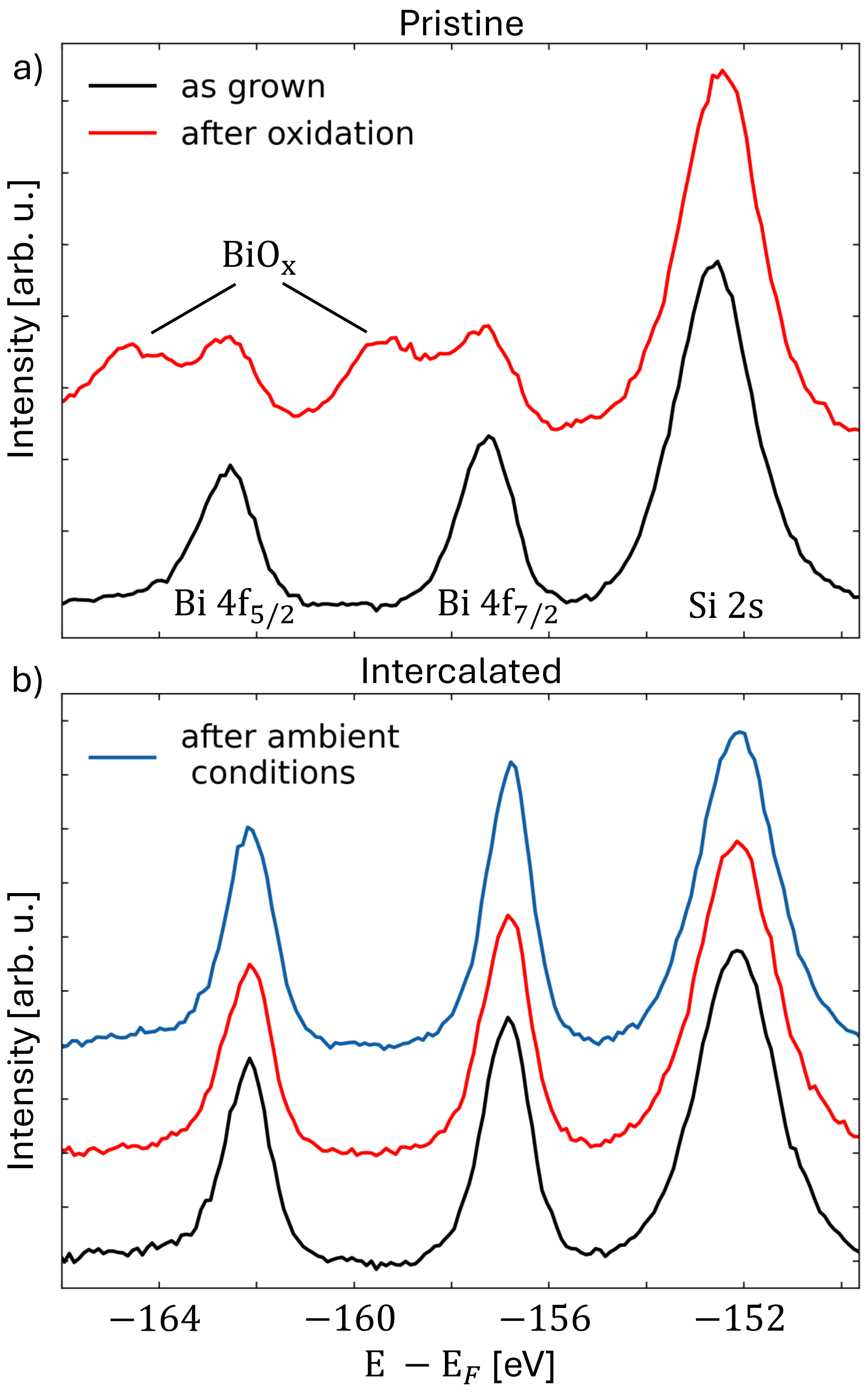}
\caption{\textbf{Resilience of intercalated bismuthene in air.} XPS spectra of the Bi $4f$ and Si $2s$ core level peaks of \textbf{a)} pristine and \textbf{b)} intercalated bismuthene measured with Al K$\alpha$ radiation. While pristine bismuthene oxidizes and exhibits XPS satellites upon exposure to oxygen, intercalated bismuthene remains stable even upon exposure to air.}
\label{fig: XPS}
\end{center}
\end{figure}


Having confirmed the topological band structure and edge states of intercalated bismuthene, let us finally turn to its robustness against oxidation in ambient conditions. To test this, we exposed both pristine and intercalated bismuthene to $\sim 2 \times 10^{-5} \, \text{mbar}$ oxygen for 30 minutes and subsequently recorded the X-ray photoelectron spectroscopy (XPS) spectra of the Bi $4f$ and Si $2s$ core levels. The background-subtracted and normalized spectra are shown in Fig. \ref{fig: XPS}a) and b), respectively. In the pristine (uncapped) sample, oxygen exposure introduces two additional Bi $4f$ satellite peaks compared to the as-grown film, indicating significant oxidation of the topmost bismuth layer. In stark contrast, the Bi $4f$ signature of intercalated (graphene-capped) bismuthene remains unchanged even after exposure to oxygen and ambient air, demonstrating the excellent protective properties of graphene in shielding the atomic QSHI from oxidation.


These results mark a significant step forward for \textit{ex situ} measurements and device fabrication involving intercalated bismuthene. The graphene cap not only protects the material but also opens up opportunities for non-destructive and cost-effective quality control methods, such as Raman spectroscopy \cite{Cedric_Raman} or other optical techniques. This could lead to new insights into exciton physics in bismuthene \cite{Raul_Excitons} and accelerate progress toward practical applications. 

{\noindent
	\textbf{Data Availability} 
The data that support the plots within this paper and other findings of this
study are available from the corresponding author upon reasonable request.
}

{\noindent
	\textbf{Acknowledgements}
We thank Felix Reis for supplying the STM image of pristine bismuthene in Fig. \ref{fig: Edge} and Gang Li for providing the DFT calculations in Fig. \ref{fig: ARPES}. We are grateful for funding support from the Deutsche Forschungsgemeinschaft
(DFG, German Research Foundation) under Germany’s Excellence Strategy through
the Würzburg-Dresden Cluster of Excellence on Complexity and Topology in Quantum
Matter ct.qmat (EXC 2147, Project ID 390858490) as well as through the
Collaborative Research Center SFB 1170 ToCoTronics (Project ID 258499086).
This research further used resources of the Advanced Light Source, which is a DOE Office of Science User Facility under contract no. DE-AC02-05CH11231. 
}

\begin{scriptsize}
\end{scriptsize} 
{\noindent
	\textbf{Author contributions}
    L.G. has realized with the help of C.S. and J.E. the epitaxial growth and surface characterization and carried out the photoelectron spectroscopy experiments and their analysis. L.G. and B.L. realized the STM measurements. T.W. performed the oxidation study for pristine bismuthene. On the experimental side, M.K. contributed the TEM images. R.C. and S.M. supervised this joint project.
}

{\noindent
	\textbf{Competing interests}
	The authors declare no competing interests.
}

%
%
\begin{scriptsize}
\end{scriptsize}
\newpage
\end{document}